\documentclass[12pt]{article}

\usepackage{iftex}
\ifPDFTeX
  \usepackage[utf8]{inputenc}
  \usepackage[T1]{fontenc}
  \usepackage{lmodern}
\else
  \usepackage{fontspec}
\fi
\usepackage[margin=2.5cm]{geometry}
\usepackage[english]{babel}
\usepackage{amsmath, amsfonts, amssymb}
\usepackage{graphicx}% Include figure files
\usepackage{dcolumn}% Align table columns on decimal point
\usepackage{bm}% bold math
\usepackage[mathlines]{lineno}% Enable numbering of text and display math
\usepackage{upgreek}
\usepackage{setspace}
\usepackage{multirow}
\usepackage{makecell}
\usepackage{rotating}
\usepackage{lipsum}
\usepackage[super,sort&compress,numbers]{natbib}
\usepackage{authblk}
\usepackage{hyperref}% add hypertext capabilities (load last)

\begin{document}
\onehalfspacing

\title{Mechanical strain induced topological phase changes of monolayer and bilayer ZrTe$_5$}

\author[1,2,3]{Zolt\'{a}n Tajkov\thanks{tajkov.zoltan@ek.hun-ren.hu}}
\author[3]{D\'{a}niel Nagy}
\author[4]{J\'{a}nos Koltai}
\author[1,2]{P\'{e}ter Nemes-Incze\thanks{nemes.incze.peter@ek.hun-ren.hu}}

\affil[1]{Hungarian Research Network, HUN-REN Centre for Energy Research, Institute of Technical Physics and Materials Science, 1121 Budapest, Hungary}
\affil[2]{MTA - HUN-REN EK Lendület "Momentum" Topology in Nanomaterials Research Group, 1121 Budapest, Hungary}
\affil[3]{ELTE E\"{o}tv\"{o}s Lor\'{a}nd University, Department of Physics of Complex Systems, 1117 Budapest, Hungary}
\affil[4]{ELTE E\"{o}tv\"{o}s Lor\'{a}nd University, Department of Biological Physics, 1117 Budapest, Hungary}

\date{\today}

\maketitle

\begin{abstract}

Two-dimensional materials offer exceptional tunability of their properties through strain, electrostatic gating, and related mechanisms.
Among these, ZrTe$_5$ stands out as a material in which band topology can be switched via mechanical deformation.
In its bulk form, it lies near a topological phase boundary, allowing efficient modulation of its topological state through strain.
While extensive studies exist on bulk ZrTe$_5$, the topological phase transitions in monolayer and bilayer forms remain largely unexplored.
Here we show that monolayer and bilayer ZrTe$_5$ exhibit distinct strain-induced topological transitions.
Mechanical deformation in monolayers can close the topological gap, while bilayers display a richer phase diagram, including topological and trivial insulating phases as well as an intermediate metallic phase.
Since thinner van der Waals materials are more amenable to strain tuning, our results identify bilayer ZrTe$_5$ as a candidate for achieving tunable topological phase transitions using moderate (1-2{\%}) strain as a tuning knob.

\end{abstract}

%\keywords{topological insulators, monolayer, DFT, phase diagram, SIESTA, VASP}

\section*{Introduction}

Since the discovery of time-reversal invariant topological insulators, achieving a controlled topological phase transition in two-dimensional materials remains an important goal in solid-state physics~\cite{murakami2007tuning,murakami2007phase,lu2017thickness,mutch2019evidence,Kovacs-Krausz2024-hu}.
Such transitions are characterized by controlled changes in the topological properties of a material's band structure~\cite{fukui2007quantum,asboth2016short}.
One effective approach to induce these transitions is by directly tuning the lattice parameters of topological materials through mechanical deformation.
ZrTe$_5$, a transition-metal pentatelluride, is located near the boundary separating weak and strong topological phases~\cite{li2016experimental,manzoni2016evidence,wu2016evidence,xiong2017three,lv2018shubnikov,mutch2019evidence,facio2022engineering,Tajkov2022-ur}, making it possible to switch between these phases using strain.
Indeed, phase transitions in bulk or thick exfoliated ZrTe$_5$ crystals have been previously observed~\cite{stillwell1989effect,niu2017electrical,lu2017thickness,mutch2019evidence,Tajkov2022-ur}, for example by the temperature tuned changes in the lattice constant~\cite{zhang2017electronic}.
In bulk ZrTe$_5$, transitions between weak and strong topological insulator phases have been predicted through ab initio calculations~\cite{Fan2017-mo} and signatures of it observed transport experiments~\cite{mutch2019evidence,Kovacs-Krausz2024-hu}, with additional evidence of a semimetal phase at the transition point measured via scanning tunneling microscopy~\cite{Tajkov2022-ur}.
Furthermore, ZrTe$_5$ exhibits several other notable phenomena, including the planar Hall effect~\cite{li2018giant}, anomalous Hall effect~\cite{liang2018anomalous}, and chiral magnetic effect~\cite{li2016chiral}.
% For bulk samples, an effective Dirac semimetal-like band structure has also been predicted based on quantum oscillation studies~\cite{liu2016zeeman}, optical conductivity measurements~\cite{chen2015optical}, and photoemission spectroscopy~\cite{li2016chiral}.

Significant effort has been devoted to and is still focused on understanding the properties of bulk ZrTe$_5$.
The bulk crystal is known to be highly sensitive to minor changes in its lattice parameters.
Transport measurements reveal non-monotonic strain dependence and a switch in the sign of the Dirac mass with deformation~\cite{mutch2019evidence}.
Existing ARPES measurements show that under external strain, the band gap increases with tensile strain and decreases with compressive strain~\cite{zhang2021observation}.
Intentionally induced, controllable mechanical distortion was shown to switch topological phases in the crystal~\cite{monserrat2019unraveling,Tajkov2022-ur}.
Whereas there has been a significant effort devoted to experimentally and theoretically exploring the topological phase changes in the bulk material, experimental studies on thin~\cite{tang2021two,Belke2021-lg} to monolayer~\cite{Zhuo2022-ae} samples of ZrTe$_5$ or its sister compound HfTe$_5$ remain limited.
Inspired by recent progress in bottom-up synthesis~\cite{Xu2024-hv} and Al$_2$O$_3$-assisted exfoliation~\cite{Zhuo2022-ae} of monolayer and bilayer ZrTe$_5$ samples, we explore their response to external mechanical deformation.
Theoretical predictions classify the monolayer as a $\mathbb{Z}_2$ time-reversal invariant topological insulator~\cite{weng2014transition}, but the response of the bilayer to mechanical deformation has not yet been explored, with only volumetric strain explored in the monolayer~\cite{weng2014transition}.
In this work, we investigate the topological phase diagram of monolayer and bilayer ZrTe$_5$ under mechanical strain, employing \textit{ab initio} methods.

Monolayer and bilayer crystals are particularly well-suited for strain tuning, as they can be integrated into polymer based flexible devices~\cite{Androulidakis2019-kf}, often in combination with other van der Waals materials to add functionality.
In such devices, it is more advantageous to use thin crystals since the strain is more easily transferred due to a smaller strain transfer length in thin crystals~\cite{Androulidakis2019-kf}.
The range of strain achievable in such setups matches the strain levels required to tune the electronic properties of ZrTe$_5$ monolayers and bilayers (up to 2{\%}).
For example, homogeneous biaxial compression of transition metal dichalcogenides such as WS$_2$ has been demonstrated up to $-1.5\%$~\cite{Cortes-Flores2025-ig}, while homogeneous biaxial tensile strains as high as $3\%$ have been reported for WSe$_2$~\cite{Hernandez-Lopez2022-dz}.
Using non-uniform deformation methods such as AFM indentation, tensile strains up to $11\%$ can be applied~\cite{Bertolazzi2011-eb}.

\section*{Results and Discussion}

\begin{figure}[ht!]
\begin{center}
	\includegraphics[width = 0.99 \textwidth]{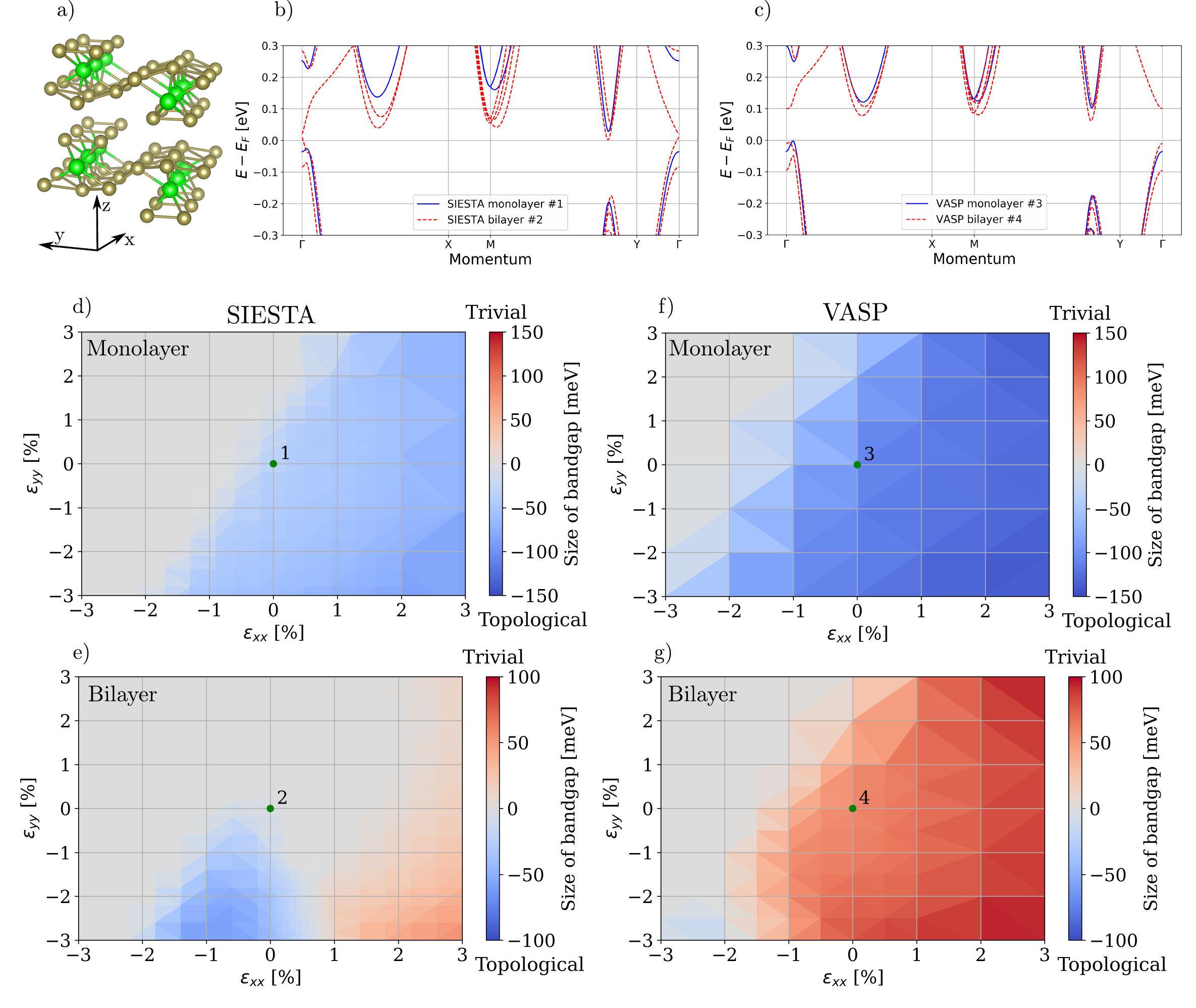}
	\caption{
		\textbf{a)} The crystal structure of the bilayer, has an orthorhombic primitive cell.
		The cell vector $b$ corresponds to the interlayer distance parallel to the $z$ direction, while vectors $a$ and $c$ lie in the $x-y$ plane~\cite{Tajkov2022-ur}.
		\textbf{b)} The band structures of the undistorted crystal obtained by SIESTA. The blue solid lines (dashed red) correspond to the monolayer (bilayer) case.
		\textbf{c)} The band structures of the undistorted crystal obtained by VASP. The blue solid lines (dashed red) correspond to the monolayer (bilayer) case.
		\textbf{d)}-\textbf{g)} The phase diagrams of the electronic structure of the crystal under mechanical strain obtained by two different DFT codes for the mono- and the bilayer cases. At each strain point, the band gap and the $\mathbb{Z}_2$ topological invariant were calculated. The color scale shows the gap magnitude with a sign used only as a plotting convention: negative values denote the $\mathbb{Z}_2$-nontrivial insulating phase, positive values denote the trivial insulating phase, and values close to zero indicate metallic or gap-closing regions.        
		}
\label{fig:1}
\end{center}
\end{figure}

Zirconium pentatelluride adopts a layered orthorhombic crystal structure.
Layers spanning the $x - y$ plane form trigonal prismatic ZrTe$_3$ lines, which are connected by zig-zag chains of Te atoms, as illustrated in Fig. \ref{fig:1}a.
The chains form two-dimensional sheets that are stacked along the $z$-axis, creating a layered structure.
The primitive cell of bulk ZrTe$_5$ contains two Zr and ten Te atoms \cite{weng2014transition}.
Each ZrTe$_5$ layer is nominally charge neutral, and the interlayer coupling is of vdW type~\cite{weng2014transition}.
The mono- and bilayer initial geometry was created by cutting one or two layers from the bulk.
The initial geometry underwent full relaxation via \textit{ab initio} calculations (see Methods section for further details).
We employed both SIESTA and VASP in our analysis of the crystal's properties to leverage the complementary strengths of each code, thus ensuring a comprehensive examination of our results.
SIESTA's localized atomic orbital basis versus VASP's plane-wave projector-augmented-wave framework allows us to cross-validate the electronic structure and topological assignment with two independent basis-set philosophies.

% The starting point for the relaxation were the lattice constants derived from our previous X-ray diffraction measurements of the bulk sample \cite{Tajkov2022-ur}.
% The relaxed and experimental geometry parameters are listed in Table \ref{table:1}.
% Our analysis reveals that the in-plane unit cell parameters and atomic coordinates within the van der Waals layers closely align with their 3D bulk experimental values, corroborating previous findings~\cite{weng2014transition}.

The initial few-layer geometries were constructed from the lattice constants obtained in our previous X-ray diffraction measurements of the bulk sample~\cite{Tajkov2022-ur}.
These experimental 300 K bulk lattice parameters were used only as structural starting information for the subsequent DFT calculations.
The relaxed DFT geometries and the experimental bulk reference parameters are listed in Table~\ref{table:1}.
The comparison is a structural consistency check between a 300 K bulk crystal and a zero-temperature few-layer DFT equilibrium structure.
Our analysis reveals that the in-plane unit cell parameters and atomic coordinates within the van der Waals layers remain close to their 3D bulk experimental values, corroborating previous findings~\cite{weng2014transition}.

The main difference between the density functional theory (DFT) codes lies in their prediction of interlayer distances ($b$) in the bilayer and the size in the $c$ unit cell vector.
In particular, the SIESTA geometry relaxations yield a $c$ distance that is 0.1 \AA\ larger and an interlayer distance that is 0.13 \AA\ less than that predicted by VASP.
Ab initio studies of bulk ZrTe$_5$ show that in-plane strain along $c$ alters the topological gap only in combination with strain along $a$ or with changes in the interlayer distance~\cite{Fan2017-mo,mutch2019evidence}, whereas the band-inversion order is controlled primarily by the interlayer spacing~\cite{manzoni2016evidence}.
From among these differences, the interlayer distance has the largest influence on the electronic bands and their predicted topological character.
Further discussion on this discrepancy will be provided later.

\begin{table}[ht!]
\begin{center}
\begin{tabular}{c| c c c c}
                        &   & $a$ [\AA] & $b$ [\AA] & $c$ [\AA] \\ \hline
Experimental                     & Bulk  & 1.994 (0.002) & 7.265 (0.005) & 13.724 (0.005) \\
\multirow{2}{*}{SIESTA} &  Monolayer & 1.999  & N.A.  & 13.870 \\
                        &  Bilayer & 1.999  & 7.172  & 13.886 \\
\multirow{2}{*}{VASP}   &  Monolayer & 2.002  & N.A.  & 13.773 \\
                        &  Bilayer & 2.001  &  7.307 & 13.758 \\  
\end{tabular}
\end{center}
\caption{Unit cell dimensions for ZrTe$_5$ as derived from the analysis of X-ray diffraction data at 300 K~\cite{Tajkov2022-ur} and from relaxed DFT calculations.
Numbers in parenthesis are one standard deviation of the measurement values.
Our X-ray diffraction determined lattice constants are in excellent agreement with previous measurements~\cite{Fjellvag1986-of}.
We note that in the few layer cases the $b$ unit cell vector represents the interlayer distance.
}
\label{table:1}
\end{table}

It has been previously predicted that a single layer of ZrTe$_5$ is a $\mathbb{Z}_2$ time-reversal invariant topological insulator, a quantum spin Hall system, with a band gap of approximately $100\, \mathrm{meV}$
 \cite{weng2014transition}.
This earlier study examined the response of the monolayer to changes in the unit-cell volume, reporting a transition from the quantum spin Hall insulator to a semimetal under volume expansion or compression. In contrast, here we map the topological phase diagram of the monolayer and bilayer as a function of the two independent in-plane strain components $\varepsilon_{xx}$ and $\varepsilon_{yy}$, which are the strains most directly accessible in flexible-substrate and van der Waals heterostructure devices.
In Fig. \ref{fig:1} b-c, we present the band structures of the unperturbed crystals, computed by both the SIESTA and VASP implementations of DFT.
The blue solid lines represent the monolayer case, while the red dashed lines correspond to the bilayer.
In the case of the monolayer, both SIESTA and VASP predict that the undistorted crystal is a topological insulator, albeit with slightly different band gaps: 110 meV for VASP and 70 meV for SIESTA.
This consensus is broken in the bilayer scenario.
% Weng et al have 100 meV, using Wien2K. They also checked with the HSE06 functional, getting a similar gap.
VASP predicts that the bilayer is a trivial insulator with a band gap of 70 meV, whereas SIESTA anticipates a metallic phase, with the gap closing at the $\Gamma$ point.
We observe that when going from monolayer to bilayers, the bands most affected by the addition of a second layer are the ones around the $\Gamma$ point, those which are also most affected by mechanical deformation~\cite{Kovacs-Krausz2024-hu,zhang2021observation}.

As previously demonstrated, the bulk ZrTe$_5$ is highly sensitive to minor deformations in the crystal structure \cite{Tajkov2022-ur}.
This sensitivity is also evident in the case of few-layer crystals.
In Fig. \ref{fig:1} d-g, we present the topological phase diagrams for mono- and bilayer ZrTe$_5$ under in-plane mechanical strain, as a function of the strain along the $x$ and $y$ directions: $\varepsilon_{xx}$ and $\varepsilon_{yy}$.
For each point of the strain phase diagrams, we calculated both the band gap and the $\mathbb{Z}_2$ topological invariant. The sign used in the color scale is a visualization convention that combines these two quantities in a single plot: negative values denote the $\mathbb{Z}_2$-nontrivial insulating phase, while positive values denote the trivial insulating phase. Values close to zero correspond to metallic or gap-closing regions. The sign should therefore not be interpreted as a physical sign of the band gap; it only indicates the topological classification associated with the calculated gap magnitude.
%For each point on the diagram, the size of the gap is calculated, and arbitrarily a sign is assigned to it in order to indicate its topological character.
%A negative gap corresponds to the topological insulating phase, whereas a positive gap signifies the trivial phase. We wish to clarify that the signs attributed to the gap values are solely for the purpose of enhancing result comprehension and do not denote any physical properties.
Our exploration involved analyzing the crystals using both SIESTA and VASP.
As illustrated in Fig. \ref{fig:1} d and f, SIESTA and VASP concur that the monolayer resides at the boundary between topological and trivial (semimetallic) phases.
Applying a modest uniaxial compression or tension, the system can be tuned from a topological insulating to metallic phase.
These results indicate that the monolayer inherits from the bulk crystal the property of residing at the boundary of a topological phase transition.

In Fig.~\ref{fig:1} e and g, we compare the topological phase diagrams of bilayer ZrTe$_5$.
SIESTA predicts that the bilayer resides at the boundary between the topological insulator and metallic phases, see green dot in Fig.~\ref{fig:1}e.
The diagram indicates that applying a moderate biaxial strain of around 1\%\ can transition the system into a topological phase, while an experimentally achievable~\cite{Androulidakis2019-kf,Cortes-Flores2025-ig,Hernandez-Lopez2022-dz,Bertolazzi2011-eb} biaxial strain can induce a trivial insulating phase through the metallic regime.

\begin{figure}[!h]
\begin{center}
	\includegraphics[width = 0.49 \textwidth]{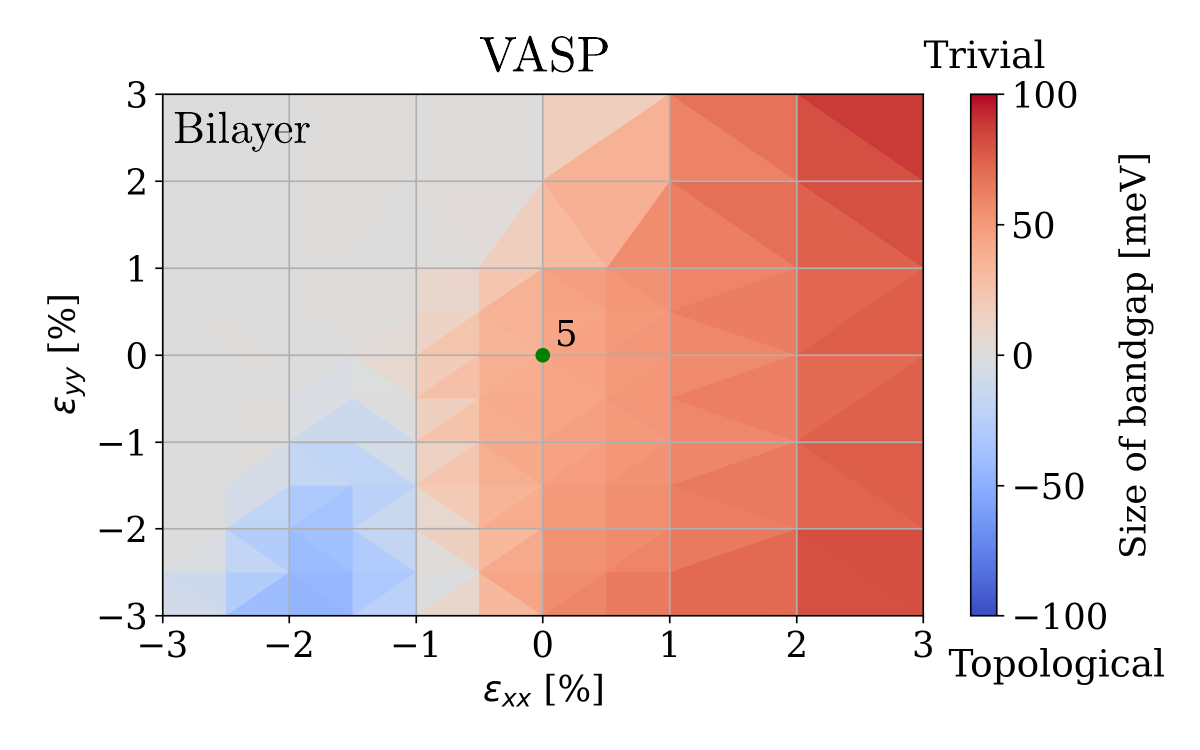}
	\caption{
		The topological phase diagram of the bilayer ZrTe$_5$ calculated by VASP. The interlayer distance has been matched to the SIESTA optimized interlayer distance.
		}
\label{fig:2}
\end{center}
\end{figure}

The differences seen in the band structures of the bilayer are also evident when comparing the phase diagrams.
In Fig. \ref{fig:1}g, it can be observed that, according to the VASP calculations, the bilayer resides further away from the phase boundary and is in a topologically trivial phase with a significant band gap of 70 meV.
However, the central feature from SIESTA, that biaxial strain can be used to transition the system into the topological phase, is confirmed by VASP, but much larger strain is required to realize the transition.
Further investigation into the role of geometry and lattice distortions clarifies the origin of the differences between VASP and SIESTA in the bilayer case.
In Fig. \ref{fig:2}, we present the topological phase diagrams of the bilayer calculated with VASP.
This time, however, the interlayer distance has been modified to align with the SIESTA optimization results; specifically, we adjusted the distance between the layers by making it 0.13 \AA\ smaller.
With this adjustment the predictions of the two codes converge substantially.
Notably, our choice to align VASP with SIESTA in this aspect was driven by our prior observations regarding the accuracy of SIESTA's predictions.
These observations, as discussed in Tajkov et al. (2022), revealed that SIESTA consistently yielded results that demonstrated better agreements with experimental data, particularly in the context of bulk properties of ZrTe$_5$ \cite{Tajkov2022-ur}. 

We interpret the bilayer transition using the monolayer band-inversion mechanism established by Weng, Dai, and Fang~\cite{weng2014transition}.
In the notation of that work, the relevant states at the $\Gamma$ point are the zigzag-chain Te$^{z}$-$p_x$ and prism-chain dimer Te$^{d}$-$p_y$ states, while the apical Te$^{a}$ states do not participate in the inversion.
As summarized schematically in Fig.~4 of Ref.~\cite{weng2014transition}, the nonsymmorphic crystal symmetry and the intra- and inter-chain hybridization invert a bonding Te$^{d}$-$p_y$ state and an antibonding Te$^{z}$-$p_x$ state.
Spin--orbit coupling then opens a gap in this already inverted manifold.
We use this established monolayer picture as the reference for identifying the corresponding mechanism in the bilayer.

\begin{figure}[ht!]
\begin{center}
    \includegraphics[width=0.99\textwidth]{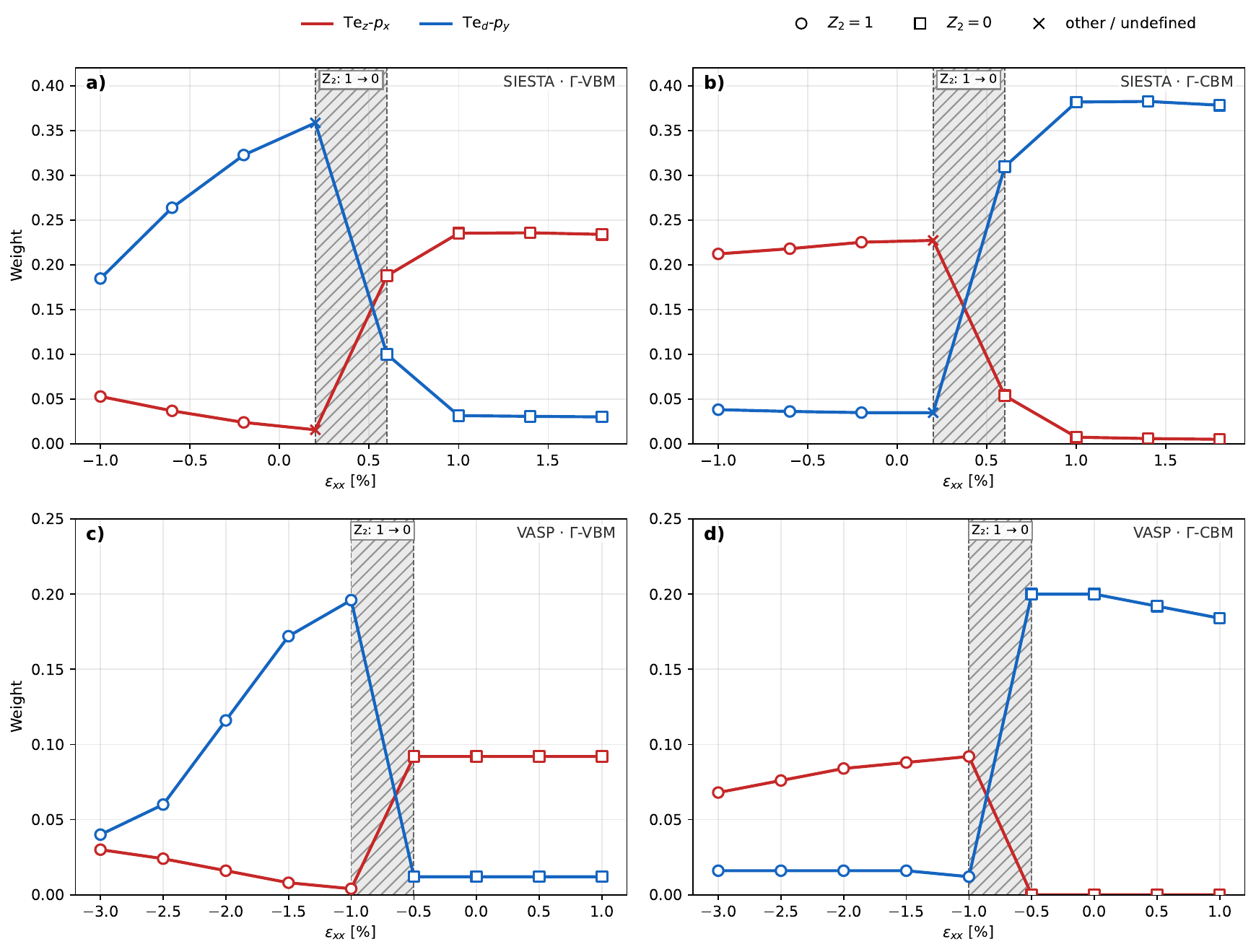}
    \caption{
    Orbital character of the $\Gamma$-point frontier states along the fixed $\varepsilon_{yy}=-1.5\%$ strain line.
    \textbf{a)} and \textbf{b)} SIESTA VBM and CBM projections, respectively.
    \textbf{c)} and \textbf{d)} VASP VBM and CBM projections, respectively.
    Red denotes the zigzag-chain Te$^{z}$-$p_x$ contribution and blue denotes the prism-chain dimer Te$^{d}$-$p_y$ contribution.
    The shaded intervals mark the strain brackets in which the $\mathbb{Z}_2$ index changes.
    The VASP panels were calculated with the interlayer distance reduced by 0.13 \AA\ to the SIESTA-relaxed value, as in Fig.~\ref{fig:2}.
    }
    \label{fig:frontier_orbitals}
\end{center}
\end{figure}

\begin{figure}[ht!]
\begin{center}
    \includegraphics[width=0.99\textwidth]{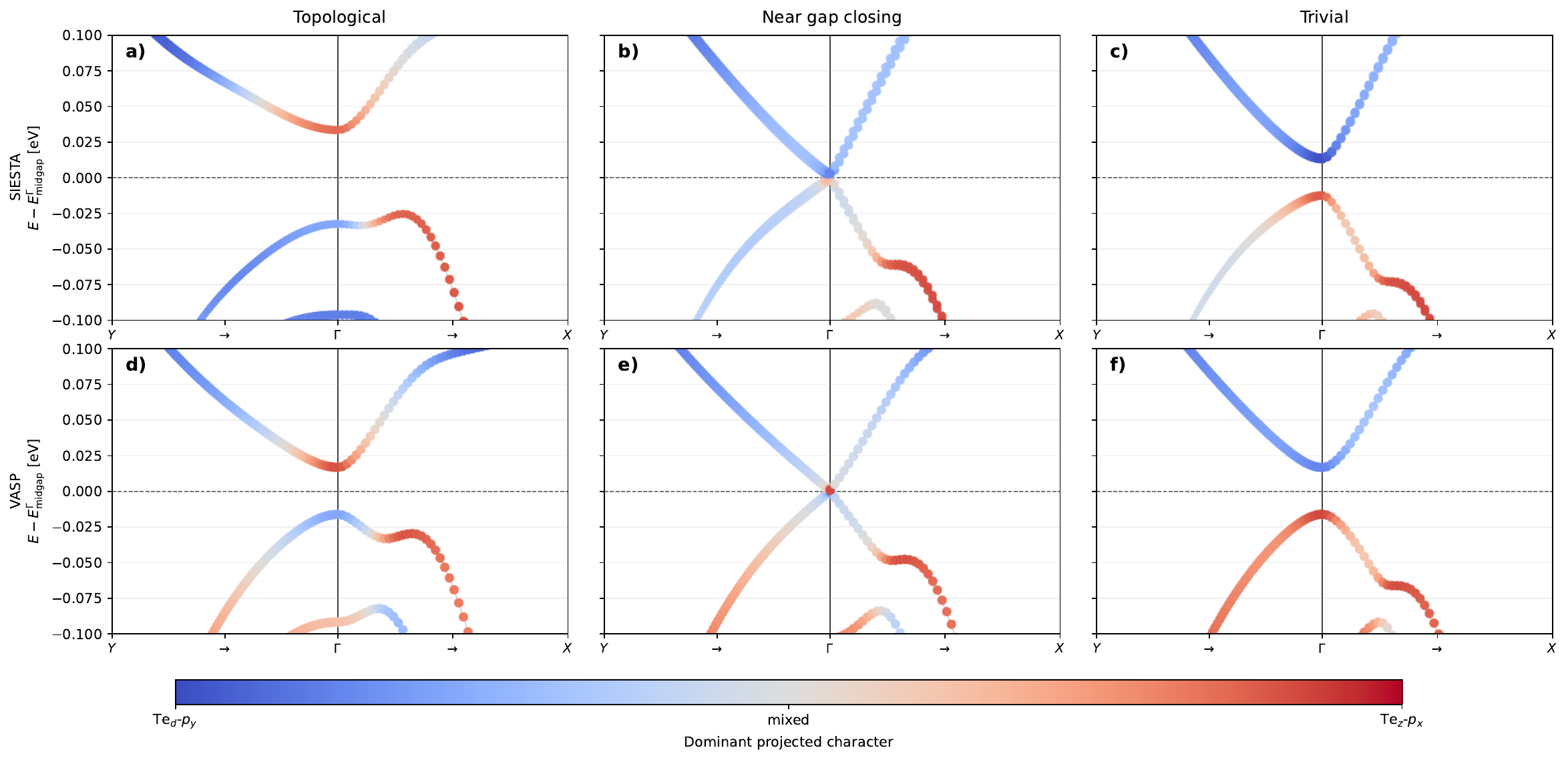}
    \caption{
    Orbital-resolved band structures near $\Gamma$ along the fixed $\varepsilon_{yy}=-1.5\%$ strain line.
    The columns show representative topological, gap-closing, and trivial configurations.
    The upper row contains the SIESTA results and the lower row the corresponding VASP results.
    The color and weight of the bands indicate the Te$^{z}$-$p_x$ and Te$^{d}$-$p_y$ contributions.
    Both methods show the closing and reopening of the direct gap together with the exchange of the dominant frontier-band characters.
    As in Fig.~\ref{fig:frontier_orbitals}, the VASP row was calculated with the interlayer distance reduced by 0.13 \AA\ to the SIESTA-relaxed value.
    }
    \label{fig:near_gamma_orbitals}
\end{center}
\end{figure}

Along the fixed $\varepsilon_{yy}=-1.5\%$ strain line, the $\Gamma$-point frontier-state projections reveal the same orbital exchange in the bilayer.
On the topological side, the VBM is predominantly Te$^{d}$-$p_y$ and the CBM is predominantly Te$^{z}$-$p_x$; on the trivial side, these dominant characters are reversed.
The exchange occurs across the gap-closing interval and coincides with the change of the $\mathbb{Z}_2$ invariant.
Figure~\ref{fig:frontier_orbitals} shows this evolution for SIESTA over $-1\%\leq\varepsilon_{xx}\leq2\%$ and for the full sampled VASP strain interval.
Here, and in Fig.~\ref{fig:near_gamma_orbitals}, the VASP calculations use the bilayer geometry with the interlayer distance reduced by 0.13 \AA\ to the SIESTA-relaxed value, as in Fig.~\ref{fig:2}.
Although the absolute projected weights differ because the two codes employ different projection schemes, both calculations give the same strain-driven exchange of the two frontier-state characters.

The projected band structures in Fig.~\ref{fig:near_gamma_orbitals} provide a complementary view of the same transition.
The three columns represent the topological phase, the vicinity of the gap closing, and the trivial phase, while the upper and lower rows show the SIESTA and VASP results, respectively.
In both calculations the direct gap closes and reopens at $\Gamma$ while the Te$^{d}$-$p_y$ and Te$^{z}$-$p_x$ characters are exchanged between the valence and conduction frontier bands.
Thus, the strain-induced bilayer transition is the bilayer counterpart of the monolayer inversion mechanism reported in Ref.~\cite{weng2014transition}, rather than a separate mechanism generated by the second layer.

\section*{Conclusions}

We mapped the strain-dependent electronic and topological phases of monolayer and bilayer ZrTe$_5$ using two independent DFT implementations.
The monolayer remains close to a topological phase boundary, whereas the bilayer supports topological-insulating, metallic or gap-closing, and trivial-insulating regimes within experimentally relevant in-plane strains.
The quantitative location of the bilayer phase boundary depends strongly on the relaxed interlayer distance, but the SIESTA and VASP phase diagrams converge substantially when this structural parameter is matched.

The orbital-resolved analysis identifies the microscopic origin of the bilayer transition.
Both the continuous frontier-state projections and the near-$\Gamma$ projected band structures show a gap closing and reopening accompanied by an exchange between the prism-chain dimer Te$^{d}$-$p_y$ and zigzag-chain Te$^{z}$-$p_x$ states.
This reproduces in the bilayer the monolayer band-inversion mechanism established in Ref.~\cite{weng2014transition} and is independently confirmed by SIESTA and VASP.
Bilayer ZrTe$_5$ therefore provides a realistic platform in which moderate mechanical strain can switch the band topology.
The calculated phase boundaries should be regarded as an intrinsic ideal-crystal reference, since substrate coupling, disorder, charge transfer, and local corrugation may shift their precise experimental positions.

\section*{Methods}
\label{methods}
The optimized geometry and electronic properties of the crystal were obtained by the SIESTA and VASP implementations of density functional theory (DFT) \cite{artacho2008siesta,soler2002siesta,garcia2020siesta,fernandez2006site,kresse_1993,kresse_1996}.

\textbf{SIESTA}

SIESTA employs norm-conserving pseudopotentials to account for the core electrons and linear combination of atomic orbitals to construct the valence states.
The generalised gradient approximation of the exchange and the correlation functional was used with Perdew Burke Ernzerhof parametrisation \cite{perdew1996generalized} and the pseudopotentials optimised by Rivero \textit{et al} \cite{rivero2015systematic} with a double-$\zeta$ polarised basis set and a realspace grid defined with an equivalent energy cutoff of 350 Ry for the relaxation phase and 900 Ry for the single-point calculations.
The Brillouin zone integration was sampled by a 30$\times$18$\times$1 Monkhorst Pack $k$-grid for both the relaxation and the single-point calculations \cite{monkhorst1976special}.
The geometry optimisations were performed until the forces were smaller than 0.1 eV nm$^{-1}$ . The choice of pseudopotentials optimised by Rivero \textit{et al} ensures that both the obtained geometrical structures and the electronic band properties are reliable.
After the successful self consistent cycles the necessary information was obtained by the sisl tool \cite{zerothi_sisl}.
The spin orbit coupling was taken into account in the single point calculations. 
The topological character of the insulating phases was determined from the $\mathbb{Z}_2$ invariant. For consistency between the two DFT implementations, the final phase labels used in the strain phase diagrams were evaluated using Z2Pack.\cite{Gresch2017-z2pack}

For the orbital analysis of the bilayer transition, the Kramers pairs immediately below and above the gap at $\Gamma$ were identified as the VBM and CBM frontier pairs.
In the non-orthogonal SIESTA atomic-orbital basis, the contribution of basis function $\mu$ to an eigenstate with coefficient vector $c$ was evaluated using the overlap-aware Mulliken expression
\begin{equation}
q_{\mu}=\mathrm{Re}\!\left[c_{\mu}^{*}(Sc)_{\mu}\right],
\end{equation}
where $S$ is the overlap.

Each state was normalized by its total generalized norm.
The Te$^{z}$-$p_x$ and Te$^{d}$-$p_y$ weights were obtained by summing the corresponding non-polarization basis functions over the selected Te atoms in both layers and over both spinor components.

\textbf{VASP}

The VASP code was used with projector-augmented wave pseudopotentials and the optB86b-vdW functional throughout our computation which has a non-local correlation correction that approximately accounts for dispersion interactions \cite{Klimes_2011}. We have found that employing this functional gives the most accurate geometries in comparison with experiments. 
The plane-wave cutoff energy was set  to $500\,\rm eV$ in all calculations. The structural optimizations were performed with $20\times 6\times 1$ Monkhorst-Pack set of $k$-points until all atomic forces fell below $3\,\rm meV/$\AA.
The band structure calculations were performed in the usual way, first the charge density was obtained in a self-consistent non-collinear calculation with a 24$\times$8$\times$1 Monkhorst-Pack set, and in the second step the band structure was evaluated along the high-symmetry lines of the Brillouin zone.
For each insulating strain point, the $\mathbb{Z}_2$ topological invariant of the VASP band structure was calculated using the Z2Pack package, which follows the evolution of hybrid Wannier charge centers. The resulting topological classification was used to assign the phase labels in the VASP strain phase diagrams.\cite{Gresch2017-z2pack}

The VASP orbital characters were extracted from the total-charge projection block of the PROCAR output.
The PAW-projected $p_x$ or $p_y$ weights were summed directly over the selected Te$^{z}$ or Te$^{d}$ atoms because the PROCAR entries are already orbital projection weights.
As for SIESTA, the results for the two $\Gamma$-point Kramers partners were averaged.
Because PROCAR PAW-projector weights and SIESTA overlap-aware Mulliken populations are different population analyses, their absolute numerical values are not expected to agree; the comparison is based on the dominant orbital character and its exchange across the transition.

In both the SIESTA and VASP workflows, for each imposed in-plane strain point, the lattice vectors were fixed according to the prescribed strain, while the internal atomic coordinates were relaxed within the corresponding strained unit cell before the electronic-structure and topological analysis was performed.

\section*{Competing interests}
The authors declare no competing interests.

\section*{Data Availability}
The datasets generated during and/or analysed during the current study are available from the corresponding author on reasonable request.

\section*{Funding}
The work was conducted within the framework of the MTA - HUN-REN EK Lendület "Momentum" Topology in Nanomaterials Research Group through project LP2024-17.
Financial support from NKFIH through grants \'{E}lvonal KKP 138144, Excellence 151372, K146156 and TKP2021-NKTA-05 are also acknowledged.
ZT acknowledges support from the J{\'a}nos Bolyai Research Scholarship of the Hungarian Academy of Sciences. This project is supported by the TRILMAX Horizon Europe consortium (Grant No. 101159646).
We acknowledge the Digital Government Development and Project Management Ltd. for awarding us access to the Komondor HPC facility based in Hungary.

\bibliographystyle{unsrtnat}
\bibliography{cikkek_mono}

\section*{Author contributions}
ZT and DN performed the calculations, under the supervision of JK.
PNI conceived and coordinated the project, with assistance from JK.
PNI and ZT wrote the manuscript with input from all authors.

\end{document}